\documentclass[prb,aps,twocolumn,nofootinbib,10pt, showpacs,superscriptaddress]{revtex4-1}
\usepackage{amssymb}
\usepackage{amsmath}
\usepackage{appendix}
\usepackage{graphicx}
\usepackage[dvipsnames]{xcolor}
\usepackage{lipsum}
\usepackage{nicefrac}
\usepackage{physics}
\usepackage{bm}
\usepackage[utf8]{inputenc}

\newcommand{\eig}[2]{\varepsilon_{#1\bm{#2}}}

\begin{document}
\title {Mixed topology ring states for Hall effect and orbital magnetism \\ in skyrmions of Weyl semimetals}

\newcommand{\PGI}{Peter Grünberg Institut and Institute for Advanced Simulation, Forschungszentrum Jülich and JARA, 52425 Jülich, Germany}
\newcommand{\RWTH}{Department of Physics, RWTH Aachen University, 52056 Aachen, Germany}
\newcommand{\mainz}{Institute of Physics, Johannes Gutenberg-University Mainz, 55099 Mainz, Germany}

\author{M. Redies}
\email{m.redies@fz-juelich.de}
\affiliation{\PGI}
\affiliation{\RWTH}

\author{F. R. Lux}
\affiliation{\PGI}
\affiliation{\RWTH}

\author{J.-P. Hanke}
\affiliation{\mainz}

\author{P. M. Buhl}
\affiliation{\mainz}

\author{S. Bl\"ugel}
\affiliation{\PGI}

\author{Y. Mokrousov}
\email{y.mokrousov@fz-juelich.de}
\affiliation{\PGI}
\affiliation{\mainz}

\begin{abstract}
As skyrmion lattices 
are attracting increasing attention owing to their properties driven by 
real-space topology,  properties of magnetic Weyl semimetals with complex $k$-space topology are drawing increasing attention. 
We consider Hall transport properties and orbital magnetism of skyrmion lattices imprinted in topological semimetals, by employing a minimal model of a 
mixed Weyl semimetal which, as a function of the magnetization direction, exhibits two Chern insulator phases separated by a Weyl state. 
We find that while the orbital magnetization is topologically robust and Hall transport properties  
exhibit a behavior 
consistent with that expected for the recently discovered chiral Hall effect~\cite{FabianPRL2020}, their evolution in the region of the Chern insulator gap is largely determined by the properties of the so-called mixed topology ring states, emerging in domain walls that separate the skyrmion core from the ferromagnetic background. In particular, we show that these localized ring states possess a robust orbital chirality which reverses sign as a function of the skyrmion radius, thereby mediating a smooth switching dynamics of the orbital magnetization.
We speculate that while the emergent ring states can possibly play a role in the physics of Majorana states, probing their properties experimentally can provide insights into the details of skyrmionic spin structures.   
\end{abstract}

\maketitle
\date{\today}

\section{Introduction}
Topological chiral spin textures such as magnetic skyrmions have established themselves as an exciting platform for the realization of novel physical effects and for innovative ideas in the realm of practical applications of magnetic systems~ ~\cite{lai_improvedtrack_2017,zhang_skyrmion-skyrmion_2015,tomasello_skyrm_strategy_2015,zhang_skyrmion-electronics_2020,zazvorka_thermal_2019,bourianoff_potential_2018,pinna_reservoir_2019, zhang_skyrmlogic_2015,luo_reconfigurable_2018,zhang_skyrmion-electronics_2020}. 
On the other side of the topology scale, magnetic materials which 
exhibit non-trivial $k$-space topology in the electronic structure $-$ such as quantum anomalous Hall insulators~\cite{deng_quantum_2020,chang_experimental_2013}, or antiferromagnetic topological insulators~\cite{otrokov_prediction_2019,niu_antiferromagnetic_2020} $-$ are at the heart of current research in solid state physics and spintronics. Bringing together the benefits of skyrmions, such as their efficient dynamics and stability, with the advantages of $k$-topological materials, normally associated with dissipationless transport and topological protection, appears to be an exciting avenue to pursue. 
To date, however, skyrmions have been mostly realized in metallic ferromagnetic materials that do not exhibit a distinct non-trivial $k$-space topology,  such as metallic ferromagnetic materials~\cite{miao_experimental_2014,munzer_skyrmion_2010}.
And while the interest in skyrmions realized in insulating materials is rising~\cite{zhang_magnonic_2020,insulator_skyrm1,insulator_skyrm2}, the emergence of global Chern insulating states in skyrmion lattices has been recently shown theoretically~\cite{lado_quantum_2015,hamamoto_quantized_2015,gobel_family_2018,gobel_antiferromagnetic_2017,gobel_magnetoelectric_2019}. 

At the same time, another flavor of $k$-space topology exhibited by magnetic Weyl semimetals is gaining increasing attention, and the number of specific material candidates which exhibit distinct topological band crossings in their electronic structure is constantly growing~\cite{wang_quantum_2017,liu_magnetic_2019}.
In three dimensions, Weyl semimetals host bands crossings known as Weyl nodes, that appear in pairs and carry an opposite non-zero topological charge. Recently, it was suggested that in two dimensions (2D) it is natural to interpret the emergence of topological band crossings in the context of {\it mixed} Weyl points, emphasizing that their topology and properties can be classified most optimally by including the magnetization direction $\hat{\mathbf{m}}$ and the mixed components of the Berry curvature tensor into the topological analysis~\cite{hanke_mixed_2017,niu_mixed_2019}. The 2D {\it mixed} Weyl semimetals thus behave in many aspects similarly to the 3D Weyl semimetals in $(\mathbf{k},\hat{\mathbf{m}})$-space. Recently, it was realized that the presence of Weyl points in spin textures in 2D or 3D Weyl semimetals can have a drastic effect on their magneto-electric properties, orbital magnetism and dynamics, see e.g. Refs.~[\onlinecite{Araki,Araki2,lux_engineering_2018,hanke_mixed_2017,niu_mixed_2019}]. 


In this work, by performing explicit  tight-binding calculations of skyrmion lattices imprinted into a 2D Weyl semimetal, we attempt to understand the role that the complex mixed topology can have on the Hall transport properties and orbital magnetism in these systems. 
Our main finding is the demonstration that properties of skyrmions of mixed Weyl semimetals, whose chemical potential resides in the vicinity of Weyl points, are largely determined by so-called ring states, which are localized at the skyrmion boundary and carry an orbital moment of a specific orbital chirality. While we discuss how the transport properties and orbital magnetization in these systems can be used to get insights into the details of spin distribution, we also consider a range of phenomena where ring states can be utilized for shaping skyrmion dynamics and for mediating the emergence of novel topological phases. 
This article is structured as follows. In Sec.~\ref{sec:model}, we describe the tight-binding model and parameters used in this paper. The details of the setup and transport calculations are given in Sec.~\ref{sec:comp_details}. In Sec.~\ref{sec:results} we present and discuss the results of our calculations, providing an outlook in Sec.~\ref{sec:discussion}.

\section{Model}
\label{sec:model}
In order to assess the transport properties of 2D skyrmions in mixed Weyl semimetals we choose the tight-binding model such that the underlying electronic structure of the ferromagnetic host exhibits a mixed Weyl point in an extended phase space of the magnetization direction and $k$-space. A similar model has been used to study the topological properties of ferromagnetic mixed Weyl semimetals in the past~\cite{hanke_mixed_2017}. The tight-binding Hamiltonian of this model on a honeycomb lattice with two structurally inequivalent atoms per unit cell and two spin-split orbitals per site reads:
\begin{equation} \label{Eq1}
   \begin{split}
    H &= \, \lambda \sum_{i \epsilon \zeta} \left( \hat{\mathbf{m}_i} \cdot \pmb{\sigma} \right) c_{i\epsilon}^\dag c_{i\zeta}\\
   -t \sum_{\left<ij\right> \epsilon} c_{i\epsilon}^\dag c_{j\epsilon}
    &+ it_{\text{so}} \sum_{\left<ij\right> \epsilon \zeta} \hat{\mathbf{e}}_z \cdot \left( \pmb{\sigma} \times \hat{\mathbf{d}}_{ij}\right) c_{i\epsilon}^\dag c_{j\zeta} 
\end{split} 
\end{equation}
The properties of {\color{black}this} model of magnetic graphene with Rashba spin-orbit interaction have been extensively studied in the past for collinear ferromagnetic and antiferromagnetic cases~\cite{castro_low-density_2008,matte_novel_2009,PhysRevB.82.161414}. In Eq.~\eqref{Eq1}, $t$ is the magnitude of the nearest neighbour hopping, $t_{\text{so}}$ is the magnitude of  Rasba-like spin-orbit coupling~\cite{bychkov_oscillatory_1984} , and $\lambda$ characterizes the magnitude of the Stoner exchange splitting~\cite{edmund_clifton_stoner_collective_1936,edmund__c.__stoner_collective_1938}. In the above expression the indices $i$ and $j$ run over nearest-neighbor atoms, while $\epsilon$ and $\zeta$ mark the spin channel. Further, $\hat{\mathbf{e}}_z$ is the unit vector in the $z$-direction (out of the two dimensional plane), while $\hat{\mathbf{d}}_{ij}$ is the unit vector connecting atom sites $i$ and $j$. The operators $c_{i\epsilon}^\dag$ and  $c_{i\epsilon}$ are the creation and annihilation operators of an electron on site $i$ with spin $\epsilon$.
The vector of Pauli matrices is denoted as $\pmb{\sigma}$, while $\hat{\mathbf{m}_i}$ is the unit vector of the magnetization direction at atom $i$, which generally varies in real space when a spin texture is present.

\begin{figure}[t!]
    \centering
    \includegraphics[width=0.47\textwidth]{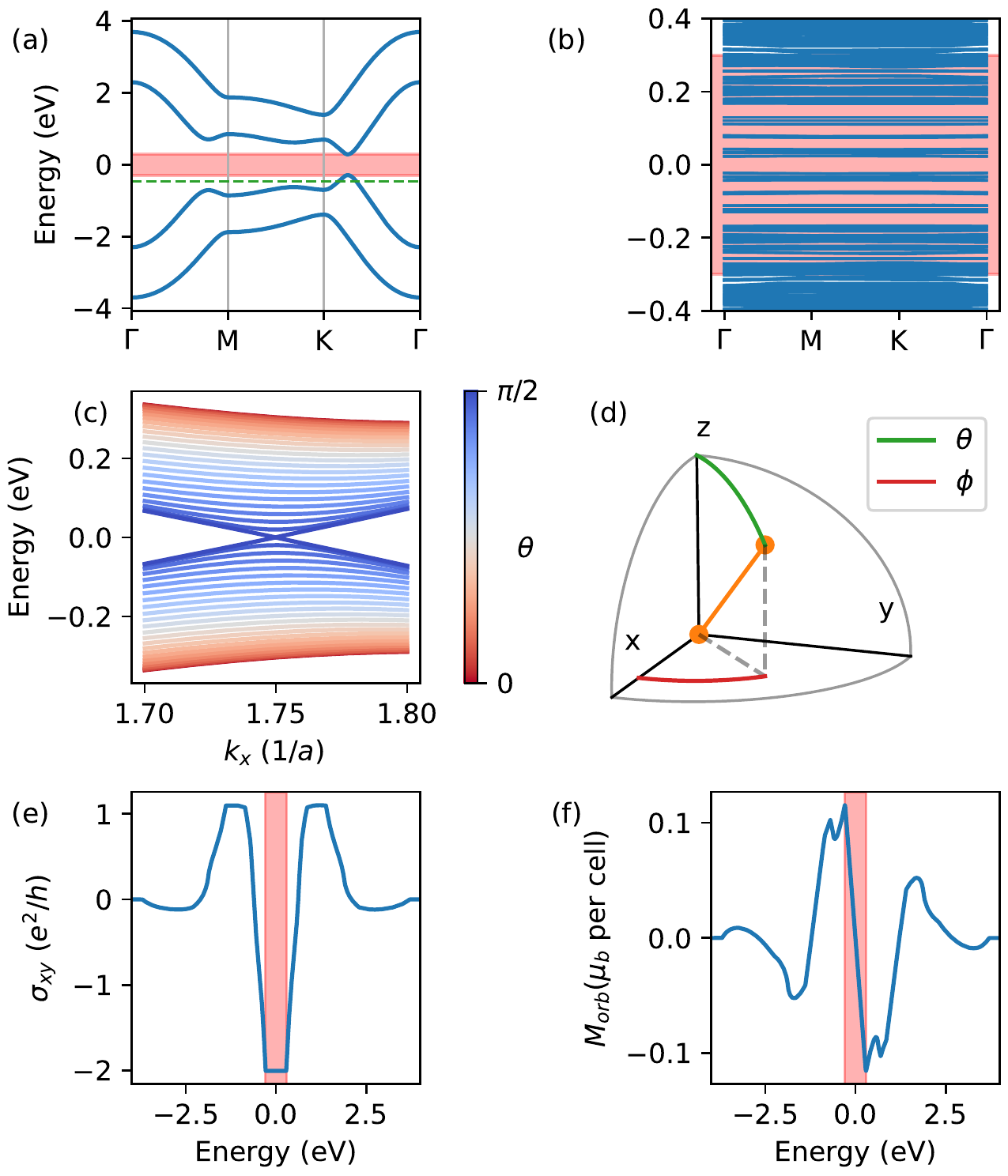}
    \caption{(a) The bandstructure of the ferromagnetic model with an out-of-plane magnetization. The green dashed line marks the region $[-0.47, -0.46]$\,eV.  (b) The bandstructure of a skyrmion with the parameterization described in the text.
    The red-shaded area in (a) and (b) indicates the band gap region of about 0.6\,eV of the ferromagnetic model. (c) A section of the bandstructure is shown for different angles $\theta$ that the magnetization in the ferromagnetic model makes with the $z$-axis. (d) The magnetization is characterized by the azimuthal angle $\theta$ and the polar angle $\phi$. The skyrmion is placed in the $xy$-plane. (e,f) The Hall conductance (e) and orbital magnetization (f) of the ferromagnetic model with an out-of-plane magnetization (along $+z$) as a function of band filling. Red shaded areas coincide with the band gap marked in (a). {\color{black} All plots except for (b) and (d) are for  $\phi=0$.}}
    \label{fig1}
\end{figure}

Throughout this work, we choose $t=-1.0$\,eV, $t_{so}=~0.4$\,eV and $\lambda = 1.4$\,eV in order to realize a mixed Weyl point in the electronic structure for the ferromagnetic case when all $\hat{\mathbf{m}_i}$ are aligned along a single direction $\hat{\mathbf{m}}$.
The bandstructure of the model for the out-of-plane direction of the ferromagnetic magnetization with these parameters is shown in Fig.~\ref{fig1}(a). The emergence of the mixed Weyl point upon varying the magnetization direction is shown in Fig.~\ref{fig1}(c). The topologically non-trivial character of this crossing point can be shown by computing the flux of the Berry curvature tensor with components in $k$- and $\theta$-space, which quantifies the mixed topological charge of the mixed Weyl point to be $+ 2$ for $\theta=\pi/2$, and $- 2$ for $\theta=-\pi/2$ ~\cite{niu_mixed_2019}.

\section{Computational approach}
\label{sec:comp_details}
Similarly to our previous work on transport properties of chiral bobbers~\cite{redies_distinct_2019} we calculate the Hall conductance and orbital magnetization of the system by employing the $k$-space Berry curvature formalism, with the only non-vanishing component of the Berry curvature tensor in $k$-space being
\begin{equation}
    \Omega_{xy}^n = -2 \Im  \braket{\frac{\partial u_{n\bm{k}}}{\partial k_x}}{\frac{\partial u_{n\bm{k}}}{\partial k_y}},
\end{equation}
where $u_{n\bm{k}}$ is the lattice-periodic part of the Bloch wave function of the band $n$ for a Bloch vector $\bm{k}$.
The intrinsic Hall conductance (in units of $e^2/h$) is then evaluated as the Brillouin zone  integral of the Berry curvature of all occupied states~\cite{iff_spring_school_2017}
\begin{equation}
    \sigma_{xy} = \sum_n^{occ} \int\limits_{\rm BZ} \Omega_{xy}^n(\bm{k})d\bm{k}.
    \label{eq:hallcond}
\end{equation}
For an insulator the Hall conductance is quantized and it is proportional to the value of an integer Chern number $\mathcal{C} := 
\int_S \Omega_{xy} dS/2\pi = 
- \sigma_{xy}/2\pi$. We calculate the out-of-plane component of the orbital magnetization (OM) in the system, $M_{orb}$,  employing the modern theory of orbital magnetism~\cite{thonhauser_orbital_2005,xiao_berry_2005,thonhauser_theory_2011}, according to which
\begin{equation}
   \begin{split}
    M_{orb} =& M^{LC}_{orb} + M^{IC}_{orb} =\frac{e}{2\hbar c}\Im \int\limits_{\rm BZ}  \frac{d\bm{k}}{(2\pi)^2}   \\
    & \sum_{n}^{occ.} 
    \mel{
        \frac{\partial u_{n\bm{k}}}{\partial k_x}
    }{
        \times \left(H_{\bm{k}} + \varepsilon_{n\bm{k}} -2\mu\right)
    }{ 
        \frac{\partial u_{n\bm{k}}}{\partial k_y}
    },
    \label{eq:orbmag}
\end{split}
\end{equation}
where $M^{LC}_{orb}$ and $M^{IC}_{orb}$ are the local circulation and itinerant circulation parts of the OM, respectively, $\varepsilon_{n\bm{k}}$ is the energy of the $n$-th Bloch state at $\bm{k}$, and $H_{\bm{k}}$ is the lattice-periodic part of the Hamiltionian $H_{\bm{k}} := e^{-i\bm{k}  \bm{r}} H e^{i \bm{k} \bm{r}}$. In order to compute the derivatives of the Bloch states in the tight-binding basis we employ first-order perturbation theory:
\begin{equation}
    \ket{\frac{\partial u_{n\bm{k}}}{\partial k_i} } = \sum_{m\neq n} \frac{\mel{u_{m\bm{k}}}{\nicefrac{\partial H_{\bm{k}}}{\partial k_i}}{u_{n\bm{k}}}}{\eig{n}{k} - \eig{m}{k} + i \eta} \ket{u_{m\bm{k}}},
    \label{eq:perturb_deriv}
\end{equation}
which results in the well-known gauge-invariant expression for the Berry curvature and OM~\cite{yao_first_2004,thonhauser_theory_2011}. In order to avoid divergences for (nearly) degenerate states a broadening $\eta = 10^{-8} \mbox{ Hartree}$ is introduced.

The calculated Hall conductance (HC) and OM of the model as a function of  band filling for the out-of-plane magnetization are shown in Fig.~\ref{fig1}(e) and (f), respectively. As evident from these plots, at half filling the system is a Chern insulator with the Chern number of $-2$. Therefore, the variation of the OM in the topologically non-trivial gap is linear with the chemical potential $\mu$, according to the relation valid for insulators~\cite{thonhauser_theory_2011}:
\begin{equation}\label{slope}
    \frac{dM_{orb}}{d\mu} = \frac{e}{(2\pi)^2\hbar c}\,\sigma_{xy},
\end{equation}
where $e$ is the electron charge and $c$ is the speed of light. The emergence of the metallic point in the spectrum of the model is due to the change in the Chern number from $-2$ to $+2$ at half filling upon reversing the direction of $\hat{\mathbf{m}}$. 

Based on the model (\ref{Eq1}) we imprint the skyrmion lattice by varying the direction of $\hat{\mathbf{m}}_i$ in real space according to the parametrization of the skyrmions discussed below.
We consider a hexagonal lattice of skyrmions where in the center of each unit cell we place a N\'eel skyrmion. {\color{black} Skyrmions of this type can be stabilized experimentally in the presence of a small external magnetic field, the effect of which on the electronic properties we do not take into account.} The chemical unit cell contains two atoms, which are located at $(0,\pm a,0)^T$. The lattice vectors of the chemical unit cell are $(l_{\text{chem}},0,0)^T$ and $l_{\text{chem}}\cdot(\frac{1}{2}, \frac{\sqrt{3}}{2}, 0)^T$, where $l_{\text{chem}} = \sqrt{3} a$ and $a$ is the distance between the inequivalent atoms in the chemical unit cell. The unit cell of the skyrmion lattice is hexagonal and contains 1568 atoms. The lattice vectors of the supercell are $(l_{\text{mag}},0,0)^T$ and $l_{\text{mag}}\cdot(\frac{1}{2},  \frac{\sqrt{3}}{2}, 0)^T$, where $l_{\text{mag}} =  28 \, l_{\text{chem}}$.
This corresponds to 28 atoms in between the centers of neighboring skyrmions and 1568 atoms in the unit cell.
We use an adaptive integration scheme to evaluate the integrals in Eqs.~(\ref{eq:hallcond}) and (\ref{eq:orbmag}). This leads to the convergence of the results presented in Figs.~\ref{fig:steepness_scan} and \ref{fig:OM_slope} with  6144 $k$-points, and results presented in Fig.~\ref{fig1}(e,f) with 16042 $k$-points. The full source code is available as open source~\cite{git_repo}.

\section{Results}
\label{sec:results}


\subsection{Emergence of mixed topology ring states in skyrmions of mixed Weyl semimetals}

Here, we analyze the electronic structure of a skyrmion lattice of a mixed Weyl semimetal at half filling, and in the following the zero of energy is associated with the position of Fermi energy for the case when exactly two electronic states per structural unit cell are occupied. For the ferromagnetic system this corresponds to the position of the metallic Weyl point in the spectrum when the magnetization is in-plane, see Fig.~\ref{fig1}(c). 
Correspondingly, in the limit of very large skyrmions, when the regions with homogeneous out-of plane magnetization in the skyrmion center and in-between the skyrmions are large, we expect an emergence of electronic states around the zero energy. This state is expected to be localized to the region where the magnetization lies in-plane, i.e. within the domain wall separating the skyrmion center with the outside region.

\begin{figure}[t!]
    \centering
    \includegraphics[width=0.49\textwidth]{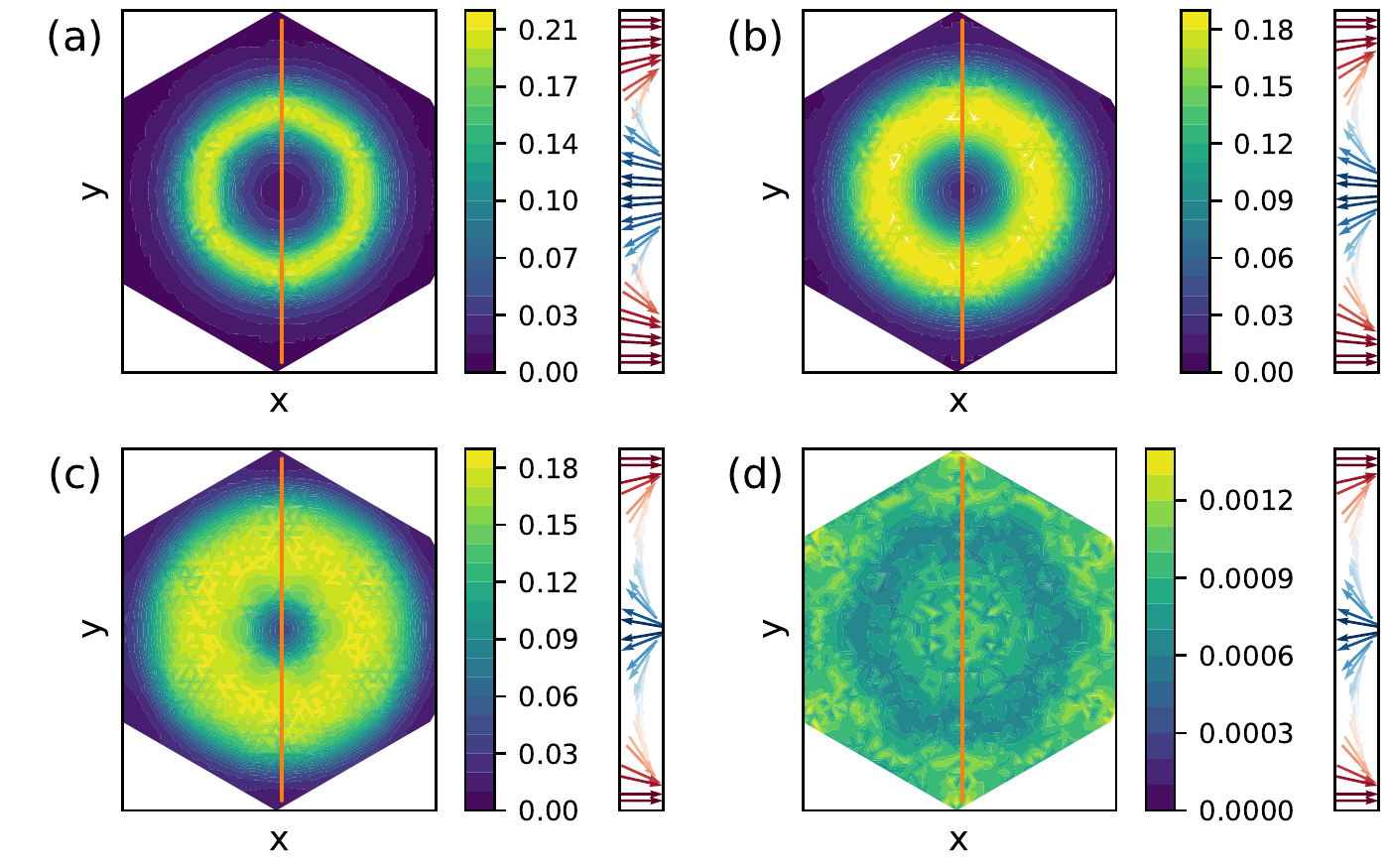}
    \caption{ Emergence of mixed topology ring states (MTRS) in a skyrmion lattice of a mixed Weyl semimetal with skyrmion radius of $r_\text{max} = \frac{1}{2} \, l_{mag}$. (a-c) The local density of states (LDOS) in real space integrated over the energy range from $-$0.3 to $+$0.3\,eV, corresponding to the gap of the ferromagnetic system in Fig.~\ref{fig1}(a,b).
    The magnitude of the integrated LDOS is indicated in the color bar. The arrows to the right indicate the spin direction along the orange line in the $(xy)$-plane. Here the dark red (blue) arrows indicate a $\mathbf{m}=(-)\hat{e}_z$ magnetization direction. In (a) $d=0.2$, in (b) $d=0.5$ and in (c) $d=0.8$. The LDOS marks the formation of MTRS. 
    (d) The LDOS  integrated in the energy range $[-0.47, -0.46] \mbox{ eV}$, away from the gap, marked green in Fig~\ref{fig1}(a). 
    The magnetic structure is identical to (c). In this energy region the MTRS are absent. In the center of the skyrmion $\mathbf{m} = + \mathbf{e}_z$, while  $\mathbf{m} = - \mathbf{e}_z$ at the edges of the unit cell.}
    \label{fig:plateau}
\end{figure}

We verify this by performing explicit calculations of the electronic structure of the skyrmion lattice with the model described above. The N\'eel skyrmions of radius $R$ (with  the maximal value of it being $R_{max}=l_{\rm mag}/2$), are parameterized by the angle $\alpha$ that a spin at the distance $r$ from the center of the skyrmion makes with the $z$-axis, while the parameter $d$ is introduced to tune the width of the domain wall region where the spins are lying in-plane {\color{black} so that larger $d$ results in a larger region of ``flat" magnetization (see examples in Fig~\ref{fig:plateau}(a-c) for different values of $d$). Note that the rate of change of the magnetization outside of the in-plane regions does not change. We use this parametrization to artificially tune the internal width of the states which emerge in the ``flat" region.} 
Below we consider the case of $R=R_{max}/2$ when introducing the parameterization in terms of $r$ and $d$ as follows: $\alpha(r, d)  =\pi  \frac{\beta(r/R_{max}, d)-\beta(1, d)}{\beta(0, d)-\beta(1, d)}$, with  $\beta(x, d)  =\operatorname{atan}\left(R_{max}(6x-3d-3)\right) +\operatorname{atan}\left(R_{max}(6x+3d-3)\right)$.
As we show below, this parameterization allows us to manipulate the spatial spread of states associated with the transition between the two out-of-plane domains.

In order to analyze the spatial localization of the electronic states, we compute the space-resolved local density of states (LDOS), presenting the results in Fig.~\ref{fig:plateau}. 
Namely, we look at the LDOS of states which are appearing in the electronic structure of the skyrmion in the gap of the out-of-plane ferromagnetic system between $-$0.3 and $+$0.3\,eV, see Fig. 1(a,b).
In Fig.~\ref{fig:plateau} (a-c) the spatial distribution of the skyrmion's LDOS, which has been integrated in this energy region, is shown for different widths of the domain wall as controlled by the parameter $d$. 
Quite remarkably, our calculations show that, irrespective of the domain wall width, the LDOS is vanishing exactly outside of the domain wall $-$  i.e., in the center of the skyrmion, and in the region between the skyrmions. This is in sharp contrast to all other states of the skyrmion lattice which lie outside of the gap in the ``metallic" region: as an example, in  Fig.~\ref{fig:plateau}(d) the LDOS of the system, which has been integrated over the energy region $[-0.47, -0.46]$\,eV (i.e. outside of the gap as marked with a green dashed line in Fig.~\ref{fig1}(a)), is finite at any point in the unit cell which marks the delocalized character of the constituting states.

The precise localization of the gap states in the domain that we observe irrespective of the domain wall width allows us to directly associate them with the emergence of the band crossing in the electronic structure of the model ferromagnetic system for the in-plane magnetization. Recently it was shown that this band crossing acquires a non-trivial topological character once the magnetization direction $\hat{\mathbf{m}}$ is included into the topological analysis. In this sense the ferromagnetic 2D model that we study is an example of what is called a {\it mixed Weyl semimetal}~\cite{hanke_mixed_2017,niu_mixed_2019} $-$ a term motivated by necessity of including the so-called mixed Berry curvature into the analysis of the band topology. Since the gap states that we observe arise as a result of the non-trivial mixed topology of our model, we refer to them as {\it mixed topology ring states} (MTRS). It has been shown, that, depending on the symmetry of the model, the mixed Weyl points can be associated with a transition between the Chern insulating phases with different Chern number arising for different directions of $\hat{\mathbf{m}}$~\cite{hanke_mixed_2017,niu_mixed_2019}. This is exactly the case for our model: as the direction of $\hat{\mathbf{m}}$ is changed from along the $z$-axis to the opposite to it, the Chern number at half-filling changes sign from $+2$ to $-2$. The skyrmion lattice that we consider thus presents a lattice of domains with an opposite Chern number separated by the domain walls, and the MTRS can be naturally interpreted as the topological edge states localized at the boundary separating the two domains.

\subsection{Hall conductance and orbital magnetization in skyrmions of a mixed Weyl semimetal}


In this section we investigate the possible influence of MTRS on the Hall transport properties and orbital magnetization exhibited by the skyrmion lattices of mixed Weyl semimetals. In order to do that, we parametrize the magnetization distribution within the skyrmion of radius $R$ in a way that $\alpha(r)= \pi \left(1- \frac{\delta(r)-\delta(0)}{\delta(R_{\text{max}}) - \delta(0)} \right)$ with $\delta(x) = -\operatorname{atan}\left(a\cdot (r - R) \right) + \frac{\pi}{2}$. Unlike the previous parameterization used in section IV.A, which exhibits a pronounced flat region in the middle of the wall, the parametrization we use from here on is designed to model most closely the skyrmions in systems which favor an out-of-plane magnetization direction, which leads to domain walls of roughly constant magnetization gradient across the wall.  
{\color{black} When compared to the approach used in the preceding subsection, the parametrization employed from now on is very close to that routinely used to model skyrmion profiles in micromagnetic studies~\cite{romming_field-dependent_2015,footnot}. It is important to note that the MTRS emerge for both types of the profile parametrization.}

\begin{figure}[t!]
    \centering
    \includegraphics[width=0.45\textwidth]{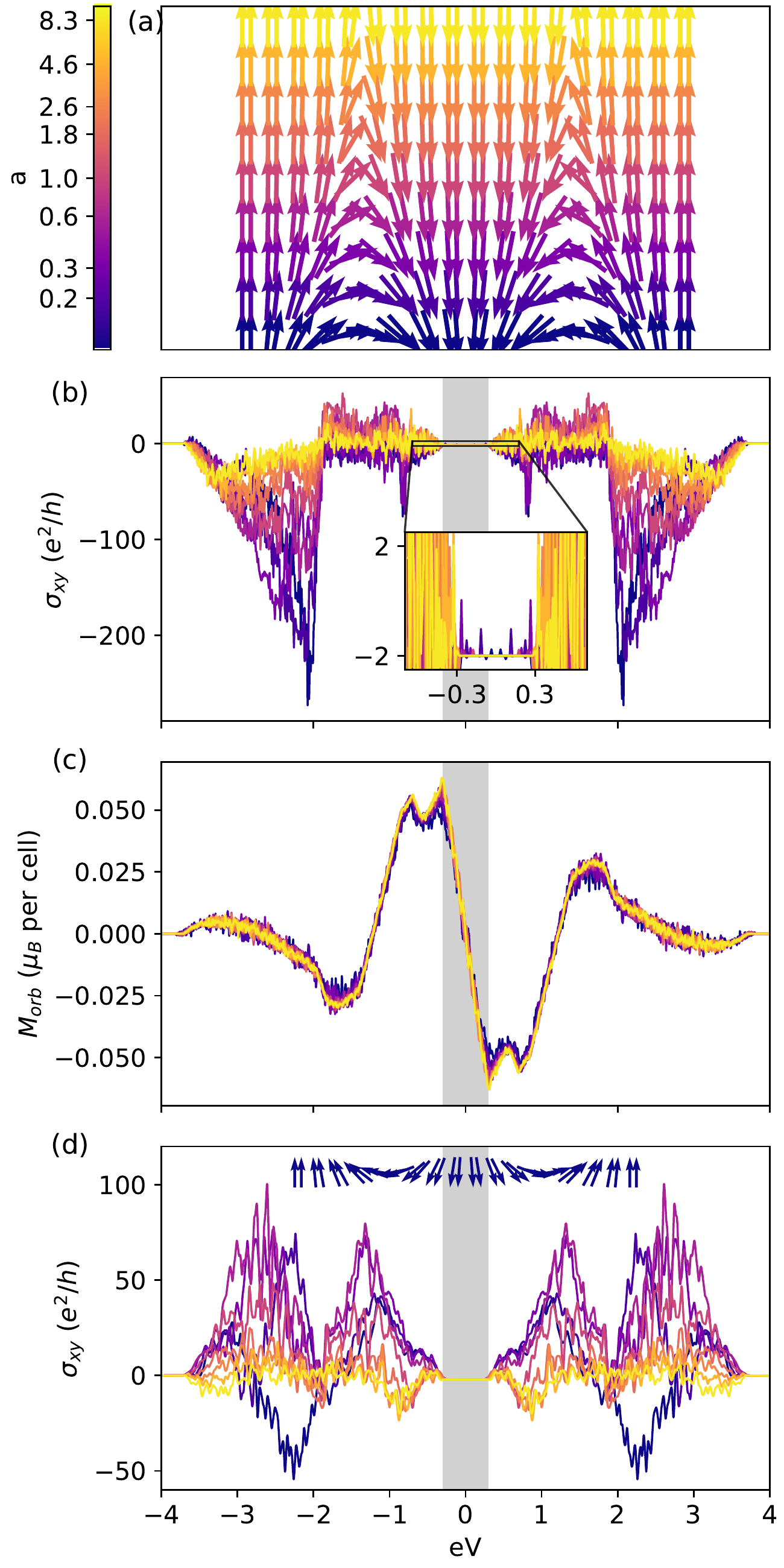}
    \caption{Transport properties of a skyrmion lattice of a mixed Weyl semimetal with $R=R_{max}/2$. 
    (a) The real-space cut through the skyrmion profile along the path connecting two skyrmion centers for different values of parameter $a$ (see text for more details). (b,c) The dependence of the Hall conductance $\sigma_{xy}$ (b), and orbital magnetization $M$ (c), on the band filling and parameter $a$. 
     The inset in (b) zooms into the region of quantization where MTRS reside.
    {\color{black} (d) Displays the HC of the skyrmions with profile shown in (a) but with reversed sense of the spin rotation in the walls (for impression see the sketch shown as an inset). }
    In all plots the color 
    indicates the value of $a$ in accordance to the color scale shown in (a).
    }
    \label{fig:steepness_scan}
\end{figure}

We first consider the case of $R = R_{max}/2$, and show the skyrmion profile along the path between the skyrmion centers for various values of $a$ in Fig.~\ref{fig:steepness_scan}(a).  
We present the results of our calculations of the Hall conductance and orbital magnetization in the system in Fig.~\ref{fig:steepness_scan}(b,c) as a function of band filling and  parameter $a$, of which the latter controls the domain wall width and correspondingly the MTRS spread in real space. Concerning the overall energy dependence for all values of $a$, while HC exhibits a symmetric structure with respect to the middle of the bulk gap at $E_F=0$\,eV, the OM is antisymmetric, which originates in the symmetry of the ferromagnetic band structure and the properties of the OM around the mixed Weyl points~\cite{niu_mixed_2019}. Overall $-$ excluding the region of the bulk gap, marked with a shaded area in Fig. 3(b,c),  which will be considered in detail below $-$ the computed HC appears to be extremely sensitive to the domain wall width. This manifestly ``non-topological'' behavior of the HC emerges in contrast to the expectations of the topological Hall effect in this system, insensitive to the details of the spin distribution but determined by the overall topological charge, with the latter remaining constant in our calculations. 

The observed sensitivity of the HC to the domain wall width,~i.e.~to the magnitude of the magnetization gradient in it, can be best understood by referring to the novel phenomenon of the {\it chiral Hall effect}~\cite{FabianPRL2020}, which has been recently shown to be prominent in chiral skyrmions with interfacial spin-orbit coupling, and which is originated in the change of the magnetization within the walls {\color{black} as given by their chirality along the line which passes through the center of the skyrmion. Within the theory of the chiral Hall effect, emerging already for spin-spiral solutions \cite{FabianPRL2020}, the chiral signal is proportional to the sense of the spin chirality among the neighboring spins $\mathbf{S}_i\times\mathbf{S}_j$, and thus the change in sign of the Hall conductivity upon changing the sign of the spin chirality serves as one of the trademarks of the chiral Hall effect~\cite{Kipp}. In Fig.~\ref{fig:steepness_scan}(d) we present the results of the HC calculations as a function of the domain wall width, but assuming that the domain wall has an opposite sense of spin rotation from that shown in Fig.~\ref{fig:steepness_scan}(a). While the topological charge of the skyrmion does not depend on the chosen chirality, the sign of the HC is opposite for two opposite chiralities for almost all values of $a$ and majority of energies. This underpins the origin of the observed HC in the chiral Hall effect originated in the domain walls of skyrmions. A perfect reversal of the HC at a given energy with chirality (i.e. reversal in sign but not in magnitude) is not expected given very strong spin-orbit interaction strength which makes the electronic structure of two flavors of skyrmions  different.} 

On the other hand, the overall value of the orbital magnetization is remarkably insensitive to the domain wall width. While the emergence of the chiral orbital magnetization,
arising in analogy to the chiral Hall effect, would be also expected for interfacial skyrmions~\cite{FabianPRL2020}, for our particular choice of the model and symmetric parametrization of the skyrmion profile the chiral part of the OM vanishes from symmetry arguments, thus reducing to the well-known effect of {\it topological orbital magnetization}~\cite{hoffmann_topological_2015,hanke_role_2016,manuel_chirality-driven_2016,hanke_prototypical_2017,lux_engineering_2018}. {\color{black} The topological OM remains extremely robust with respect to the changes in the details of spin distribution, as visible in  Fig.~\ref{fig:steepness_scan}(c), and its behavior is manifestly not very sensitive to the change in the spin chirality of the skyrmion (not shown)}.


In the following, we focus on the region in energy which corresponds to the position of the bulk gap of the model for the out-of-plane magnetization, marked with a shaded area in Fig.~\ref{fig:steepness_scan}(b-c), and zoomed into in the inset of Fig.~\ref{fig:steepness_scan}(b). The behavior of the HC, shown in the inset, displays isolated spikes on the background of a plateau value of $-2$\,$\frac{e^2}{h}$. 
{\color{black} The latter value can be traced back to the quantized value of the HC for the out-of-plane ferromagnetic model, see Fig.~\ref{fig1}(e), as for the given value of $R$
the overall out-of-plane magnetization in the unit cell is positive.}
On the other hand, the spikes at isolated positions correspond to the contributions of ring states to the HC. These contributions become somewhat more pronounced as the width of the domain wall increases, owing to the increased spread of the ring states and increased probability of electron hopping among the ring states positioned in different unit cells.
In order to investigate this effect in more detail, we study the dependence of the HC and OM on the radius of the skyrmion $R$, while assuming a very large value of parameter $a$, thus localizing the ring states in a narrow region of the domain wall, see sketches in the right column of Fig.~\ref{fig:bandstr_OM}.
The overall qualitative behavior of the HC and OM with $R$, discussed below, does not qualitatively depend on the value of parameter $a$ chosen.

\begin{figure}[t!]
    \centering
    \includegraphics[width=0.43\textwidth]{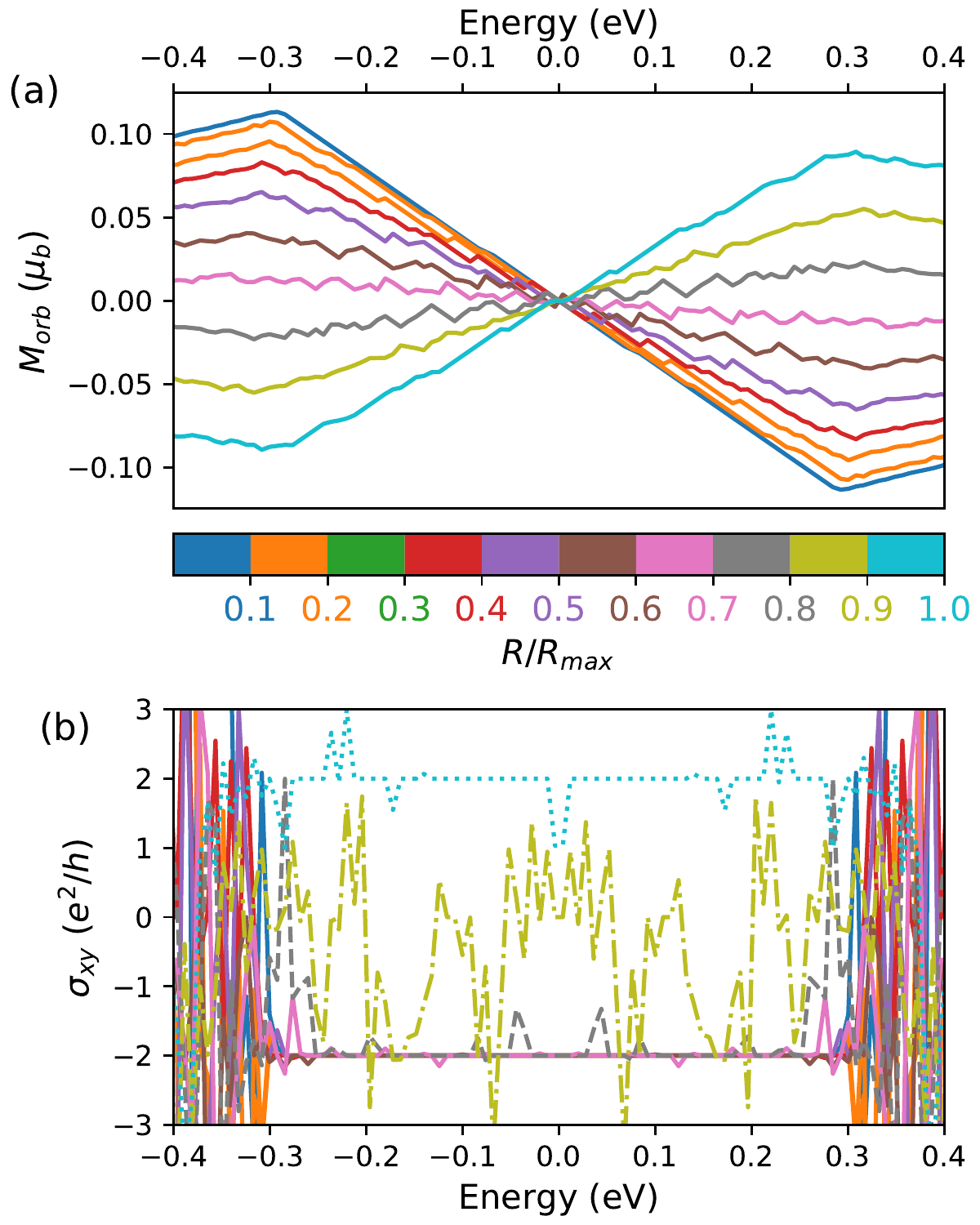}
    \caption{Transport properties of a skyrmion lattice of a mixed Weyl semimetal in the region of the bulk gap as a function of relative radius $R/R_{max}$ and a value of parameter $a$ of 10$^5$. (a) The energy dependence of the orbital magnetization as a function of $R/R_{max}$. The color of the line corresponds to the magnitude of the relative radius according to the color bar. (b) The energy dependence of the Hall conductance as a function of $R/R_{max}$. The HC of the systems with $R/R_{max} \leq 0.7$ are plotted with a solid line ``$-$", with $R/R_{max} = 0.8$ is plotted with a dashed line ``$---$",  with $R/R_{max} = 0.9$ is plotted with a dashed-dotted line ``$-\cdot-$", and with $R/R_{max} = 1$ is plotted with a dotted line ``$\cdot \cdot\cdot $". } 
    \label{fig:OM_slope}
\end{figure}

\begin{figure}[t!]
\begin{center}
    \includegraphics[width=0.47\textwidth]{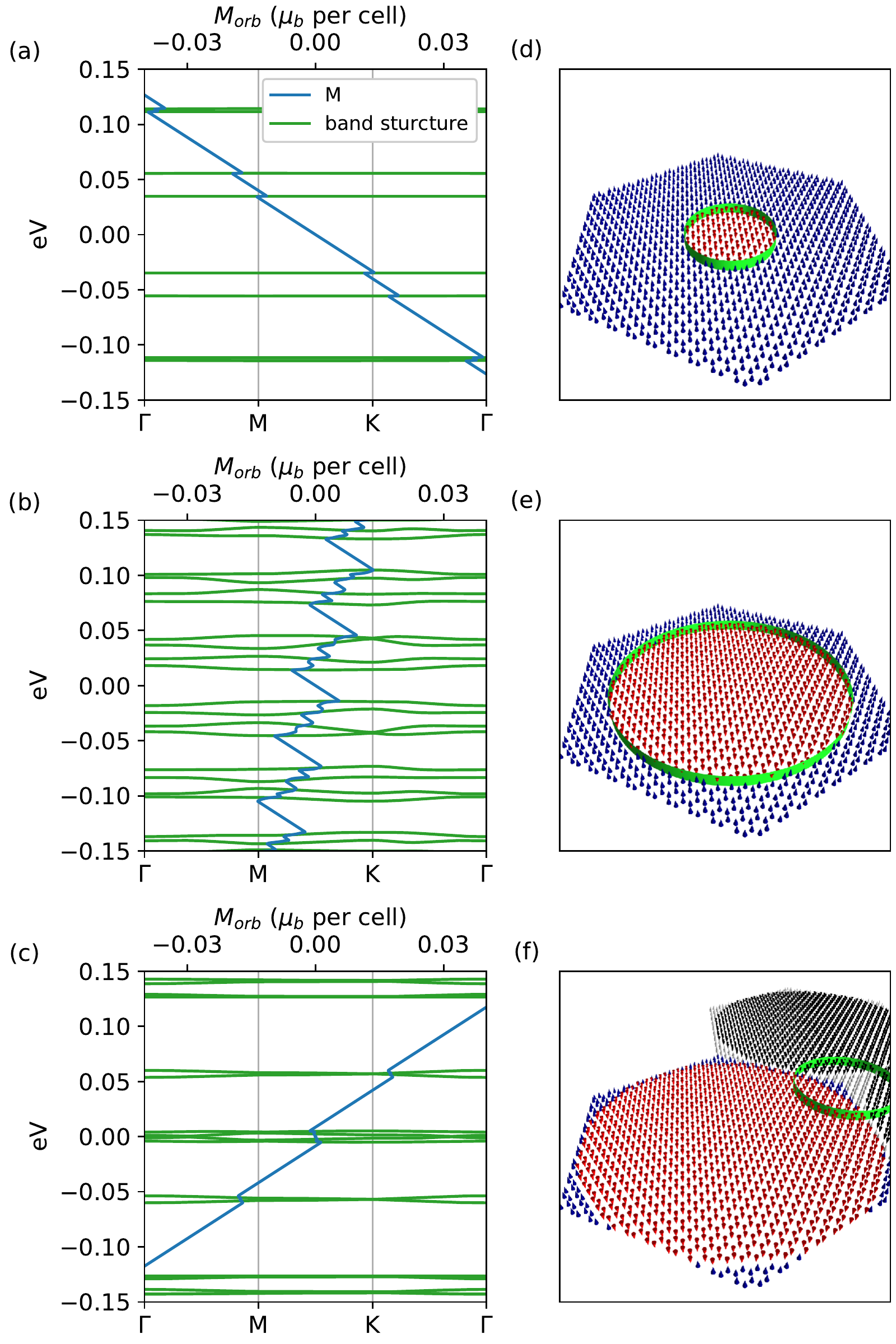}
    \caption{The evolution of the bandstructure around the Fermi energy in the region of MTRS (in green, scale on the left) and of the 
    orbital magnetization $M$ (in blue, scale on top, as a function of band filling) with respect to $R/R_{max}$.
    The positions of the jumps in the saw-tooth pattern of the orbital magnetization coincide with the positions of the MTRSs.  The sketches on the left mark the
    division of the unit cell into regions with roughly magnetization up (tainted in blue), and roughly magnetization down (tainted in red), with  $R/R_{max} = 0.3$ in (d), in $R/R_{max} = 0.8$ in (e), and $R/R_{max} = 1$ in (f). }
    \label{fig:bandstr_OM}
    \end{center}
\end{figure}

We first discuss the evolution of the HC with the relative radius $R/R_{max}$, presented in Fig.~\ref{fig:OM_slope}(b). 
Over a very large range of radii $R<0.8R_{max}$ the conductance is ideally quantized to $-2$\,$\frac{e^2}{h}$ in the gap, despite a large number of MTRS, see Fig.~\ref{fig:bandstr_OM}(a-b). In this case the Hall effect is dominated by the ferromagnetic background in between the skyrmions with $\hat{\mathbf{m}}$ along $+z$ which provides a Chern number of $-2$. The ring states are strongly localized in the wall region (which explains the flat dispersion of the ring states visible in Fig.~\ref{fig:bandstr_OM}(a)), do not hybridize with each other across the unit cell boundaries, and do not contribute to the HC. In the other limit of $R=R_{max}$, see Fig.~\ref{fig:bandstr_OM}(f) for a sketch, the film comprises mainly areas of the texture with $\hat{\mathbf{m}}$ along $-z$,
which explains mostly quantized value of the HC of $+2$\,$\frac{e^2}{h}$. {\color{black} The crossover point of the HC from $\sigma_{xy} = -2\,\frac{e^2}{h}$  to $\sigma_{xy} = 2\,\frac{e^2}{h}$  positioned between $R/R_{\text{max}} = 0.7$ and $R/R_{\text{max}} = 0.8$ coincides with the point where the integrated over the unit cell out-of-plane component of the magnetization changes sign as a result of the competition between  domains with $\hat{\mathbf{m}}$ along $+z$ and domains with $\hat{\mathbf{m}}$ along $-z$.}
The region of $0.8R_{max}\leq R \leq R_{max}$ is the region of where the MTRS interact with each other strongly {\it across} the unit cell boundaries, exhibit a strong dispersion (see Fig.~\ref{fig:bandstr_OM}(b)) and give rise to a drastic variation in the HC. In the limit of $R=R_{max}$ the MTRS are localized in the unit cell corners, thus interacting noticeably {\it along} the unit cell boundaries, which gives rise to a small but finite dispersion of the bands (see Fig.~\ref{fig:bandstr_OM}(c)) and peak-like contributions  to the HC.

We come now to the discussion of the OM of the skyrmion lattice exhibited in response to the variation of the radius. This behavior, presented in Fig.~\ref{fig:OM_slope}(a), is distinctly different from that exhibited by the Hall conductance. Clearly, in the limiting cases of $R=0$ and $R=R_{max}$ the slope of the OM as a function of energy in the gap has to be opposite, as follows from Eq.~(\ref{slope}), however, while in the case of the HC the transition between the two Chern insulator cases is relatively abrupt, the same transition in terms of the OM seems to be quite smooth, with the overall slope of the OM changing continuously as a function of $R$, and the OM curves being quite smooth as a function of energy, when compared to the case of the HC. This can be understood by looking in detail at the contributions of the MTRS to the OM variation in the gap, Fig.~\ref{fig:bandstr_OM}(a-c). Here, we observe that although the slope of the OM in between the groups of MTRS is strictly constant and consistent with the sign of the quantized value of the HC in the same region, Fig.~\ref{fig:bandstr_OM}(b), the overall curve of the OM for all values of $R$ displays a characteristic saw-tooth shape, where the overall slope is ``renormalized'' by the contributions from MTRS. Interestingly, these contributions are consistent in sign among all MTRS for a given $R$: they are positive in the region of $R<0.8R_{max}$, and negative for $R\geq 0.8R_{max}$. This ``coherence" of MTRS is the reason that the slope of OM is getting consistenly and gradually increased as $R$ goes from zero to $R_{max}$. 

According to our calculations, the OM of MTRS is predominantly originated in the local circulation part, which is consistent with the strong localization of the ring states. In a sense, the ring states can be envisaged as a group of localized electronic states which have a common unique sense of bound orbital current circulation around the skyrmion center $-$    which we can naturally refer to as having a specific sense of {\it orbital chirality}. And while in the case of the Hall effect it is the Hall conductance which changes abruptly at the point of phase transition between two areas of out-of-plane magnetized domains, the corresponding marker in terms of orbital magnetism is the orbital chirality of the ring states which exhibits a sharp transition. {\color{black} Manifestly, as follows from our calculations of the OM with reversed spin chirality of skyrmions, while the number and exact energetic positions of the states depend on the width of the domain wall for both spin chiralities, the orbital chirality of MTRS is remains a robust quantity even under the change in the skyrmion chirality.}

\section{Discussion}
\label{sec:discussion}
In our work, we have considered the intrinsic Hall effect and orbital magnetism exhibited by skyrmion lattices of mixed Weyl semimetals by referring to the Berry phase framework. Our main finding is the discovery of the special type of ring states which form as a result of complex mixed topology in real-space textures of these materials. These electronic states reside at the domain wall region, which serves as the boundary between the two Chern insulator phases and separates the skyrmion cores from the ferromagnetic background. We have shown that the degree of the localization of the ring states and the strength of their inter-cell interaction can be controlled by the parameters which determine the texture details. In turn, the ring states mediate the transition between the two Chern insulating phases occurring upon switching of the magnetization direction from out-of-plane to the opposite, when e.g. an external magnetic field changes its strength and sign in experiment. This concerns the Hall conductance, but mostly the orbital magnetization of the samples, as the ring states carry a specific orbital chirality and orbital momentum. 

The scaling of the Hall conductance as a function of the skyrmions domain wall width is a strong indicator of a prominent chiral Hall effect, where such a behavior has recently been predicted and which could have possible ramifications for the unambiguous electrical detection of magnetic skyrmions~\cite{FabianPRL2020}. In simplified model systems, the emergence of this effect can be understood from the interplay of spin-orbit interaction and exchange coupling which gives rise to an effective magnetic field~\cite{nakabayashi_rashba-induced_2014}. This phenomenology is similar to the topological Hall effect (THE)~\cite{bruno_topological_2004}, but in contrast, it materializes already at the leading order perturbation theory whereas the THE is subleading~\cite{FabianPRL2020}. From the viewpoint of Weyl semimetals, the coaction of chiral magnetism and spin-orbit coupling (which is responsible for the chiral Hall effect) is directly linked to an emergent chiral anomaly~\cite{ilan_pseudo-electromagnetic_2020}. Our model elevates this phenomenological interpretation to a more realistic setting. To which degree this explanation can be uphold will be the subject of future investigations.

Ultimately, the ring states arise as a result of a strong variation in the electronic structure around a specific magnetization direction, and thus it is rewarding to explore in the future what impact can the ring states have on the local real-space variation of the spin-orbit torques and the Dzyaloshinskii-Moriya interaction, as well as current-induced skyrmion dynamics~\cite{hanke_mixed_2017,niu_mixed_2019,Araki,hanke_engineering_2020,liu_skyrmion_2013,111A,PhysRevB.100.134407}, where the ring states can play a role of local ``hooks" strongly coupling skyrmions to applied currents. On the other hand, the unique orbital properties of the ring states lend themselves as a possible platform for obtaining a detailed information on the skyrmion parameters with e.g. XMCD type of techniques. Additionally, peculiar topological nature of these one-dimensional edge states emerging in skyrmions of Weyl semimetals hints at exceptional possibilities that the ring states can play in mediating the physics of Majorana states emerging upon deposition of skyrmions on superconductors~\cite{pershoguba_skyrmion-induced_2016,gungordu_stabilization_2018,palacio-morales_atomic-scale_2019,yang_majorana_2016,rex_majorana_2019,palacio-morales_atomic-scale_2019}, which presents an exciting future direction to pursue.

\section{Acknowledgements}
We gratefully acknowledge computing time on the supercomputers JUQUEEN and JURECA at Jülich Supercomputing Center,  and at the JARA-HPC cluster of RWTH Aachen. We  acknowledge  funding  under SPP 2137 ``Skyrmionics" of  Deutsche  Forschungsgemeinschaft (DFG), DARPA TEE program through grant MIPR\# HR0011831554 from DOI. We gratefully acknowledge financial support from the European Research Council (ERC) under the European Union's Horizon 2020 research and innovation program (Grant No. 856538, project "3D MAGiC”).

\bibliography{MTRS}
\end{document}